\documentclass{aastex631}

\usepackage{subfigure}
\usepackage{natbib}

\begin{document}

\title{ Assessing the Accuracy of TESS Asteroseismology with APOGEE }

\author[0009-0000-7811-0726]{Artemis Theano Theodoridis}
\affiliation{Department of Astronomy, University of Florida, USA }

\author[0000-0002-4818-7885]{Jamie Tayar}
\affiliation{Department of Astronomy, University of Florida,  USA }

\begin{abstract}

The recent NASA TESS mission has the potential to increase the available asteroseismic sample dramatically, but its precision and accuracy have yet to be confirmed. To date, NASA's Kepler mission has been considered the gold standard for asteroseismic samples, despite data only being available for a small portion of the sky. TESS’s observations cover the whole sky, and previous work has identified 158,000 potential red giant oscillators. Using APOGEE, which is calibrated to the asteroseismic scale of the Kepler data, we show that seismology from TESS is calibrated to the Kepler scale to better than 5\% for about 90\% of red giants, and has only slight trends with mass, metallicity, and surface gravity. We therefore conclude that current TESS seismic results can already be used for galactic archaeology, and future results are likely to be highly transformational to our understanding.

\end{abstract}

\section{Introduction} \label{sec:intro}
There has been much work done to understand the history and evolution of our Milky Way galaxy. Despite these successes, there remains a need for exact stellar ages for the individual single red giants. However, most standard methods, including measuring isochrones and analyzing chemistry, are not sufficiently precise. Asteroseismology, the study of oscillations, has shown promise in providing both precise and accurate stellar ages \citep{Zinn2022}. 
Unfortunately, previous analyses have been limited to small fields in the sky. With TESS, an all-sky survey including millions of red giants, and at least 158,000 identified oscillators, limited sky coverage is no longer a great concern \citep{Hon2021}. As exciting as this survey is, it is unclear how accurate its measurements are due to the shorter data collection period and the inclusion of machine learning in the seismic identification. Therefore, it is imperative to review these previously unconfirmed measurements. To validate its accuracy, it must be compared to the seismic scale using data that has already been thoroughly reviewed. We use spectroscopic results from the APOGEE survey for this purpose.

\section{Methods} \label{sec:style}
	We refer to the APOGEE (Apache Point Observatory Galactic Evolution Experiment) \citep{Majewski2017} Data Release 17 \citep{Abdurro'uf2022}, of the Sloan Digital Sky Survey \citep{Blanton2017} project, whose purpose is to identify an archaeological record of the Milky Way galaxy through the collection of chemical abundances and radial velocities \citep{Santana2021}. APOGEE collected near-infrared spectra using the APOGEE spectrographs \citep{Wilson_2019}. The spectra were processed with an automated pipeline  \citep{Garcia2016} using Synspec atmosphere grids. Results were calibrated using previous asteroseismic results, open clusters, and low extinction fields \citep{Jonsson2020}, and stars with poor results were flagged or eliminated. 
 The data we use from TESS comes from the asteroseismic analysis of \citet{Hon2021}. That study used machine learning to analyze long-cadence photometry from TESS taken at 30-minute intervals for one month's duration (27 days). That analysis identified potential oscillations in 158,000 red giant stars. For comparison, previous work has identified $\sim$20,000 red giant oscillators from Kepler (4-year duration), $\sim$20,000 red giants from K2 ($\sim$70-day duration), and 1800 Red giants in CoRot fields ($\sim$1-3 month duration). 
 
 Therefore, the TESS sample represents a significant potential increase in the available targets compared to previous work. 
Using the asteroseismic results, and the spectroscopic data, we created a file containing values from TESS and those from APOGEE in a similar fashion to APOKASC. Specifically, we used the correspondence between TIC\_ID and 2MASSID from TIC-v7 \citep{Stassun2019}, eliminating any stars that did not have a matching counterpart in the other dataset.  Following this process, we found 15018 matches. We added to our table calculations of seismic log(g) and mass in a similar fashion to \citet{Hon2021}, using GAIA DR2 radius \citep{Gaia2018}, APOGEE DR17 temperature, and $\nu_{\rm max}$. Once we calculated the seismic log(g) and mass, we used standard equations for error propagation to calculate errors for both seismic log(g) and mass. The catalog can be viewed at this link: \href{https://zenodo.org/record/7814297} {https://zenodo.org/record/7814297} 


\begin{figure}[tb]
\begin{center}
\subfigure{\includegraphics[width=8.5cm,clip=true, trim=0.0in 0.0in 0in 0.4in]{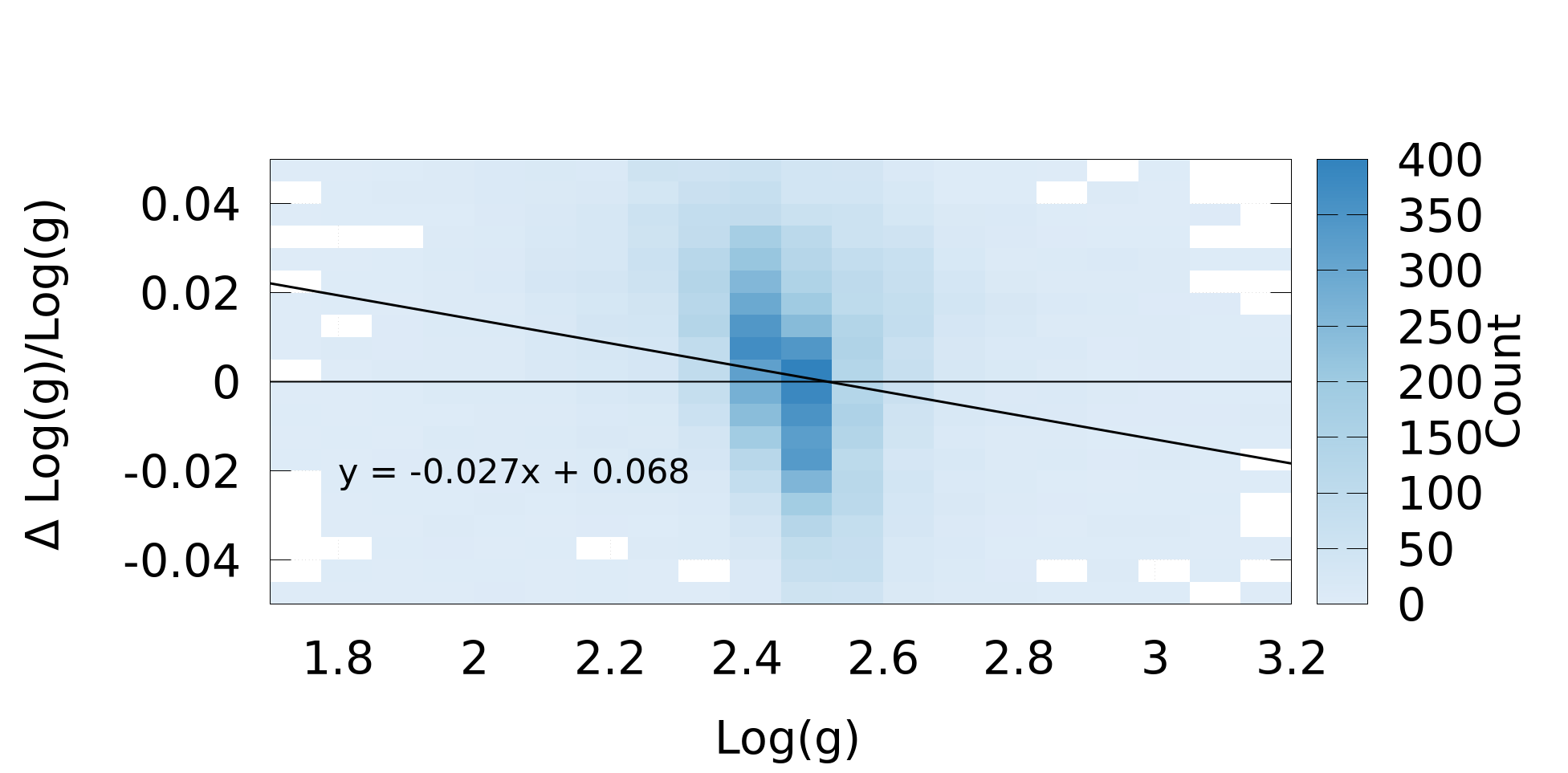}}
\subfigure{\includegraphics[width=8.5cm,clip=true, trim=0.0in 0.0in 0in 0.4in]{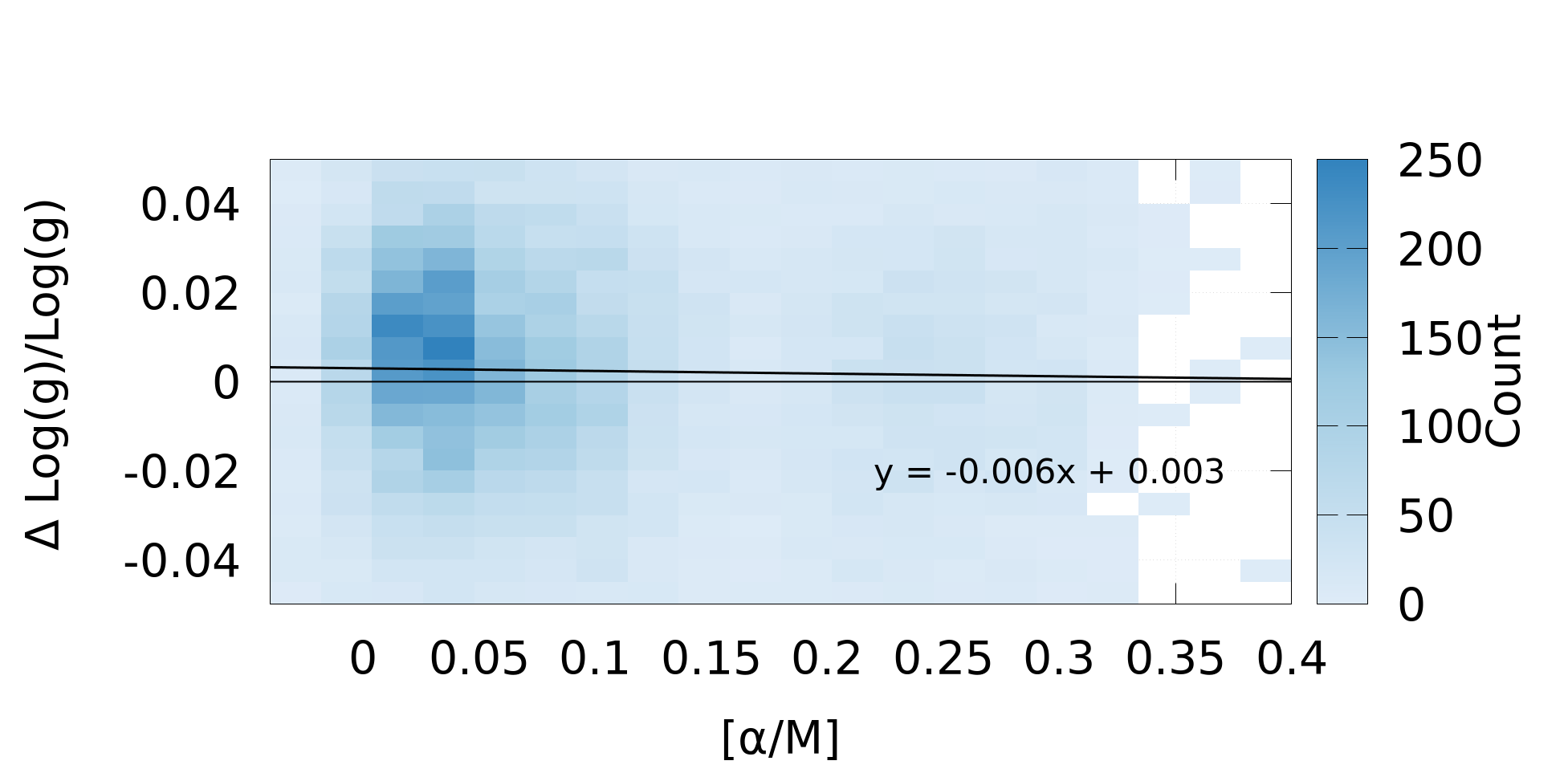}}
\subfigure{\includegraphics[width=8.5cm,clip=true, trim=0.0in 0.0in 0in 0.4in]{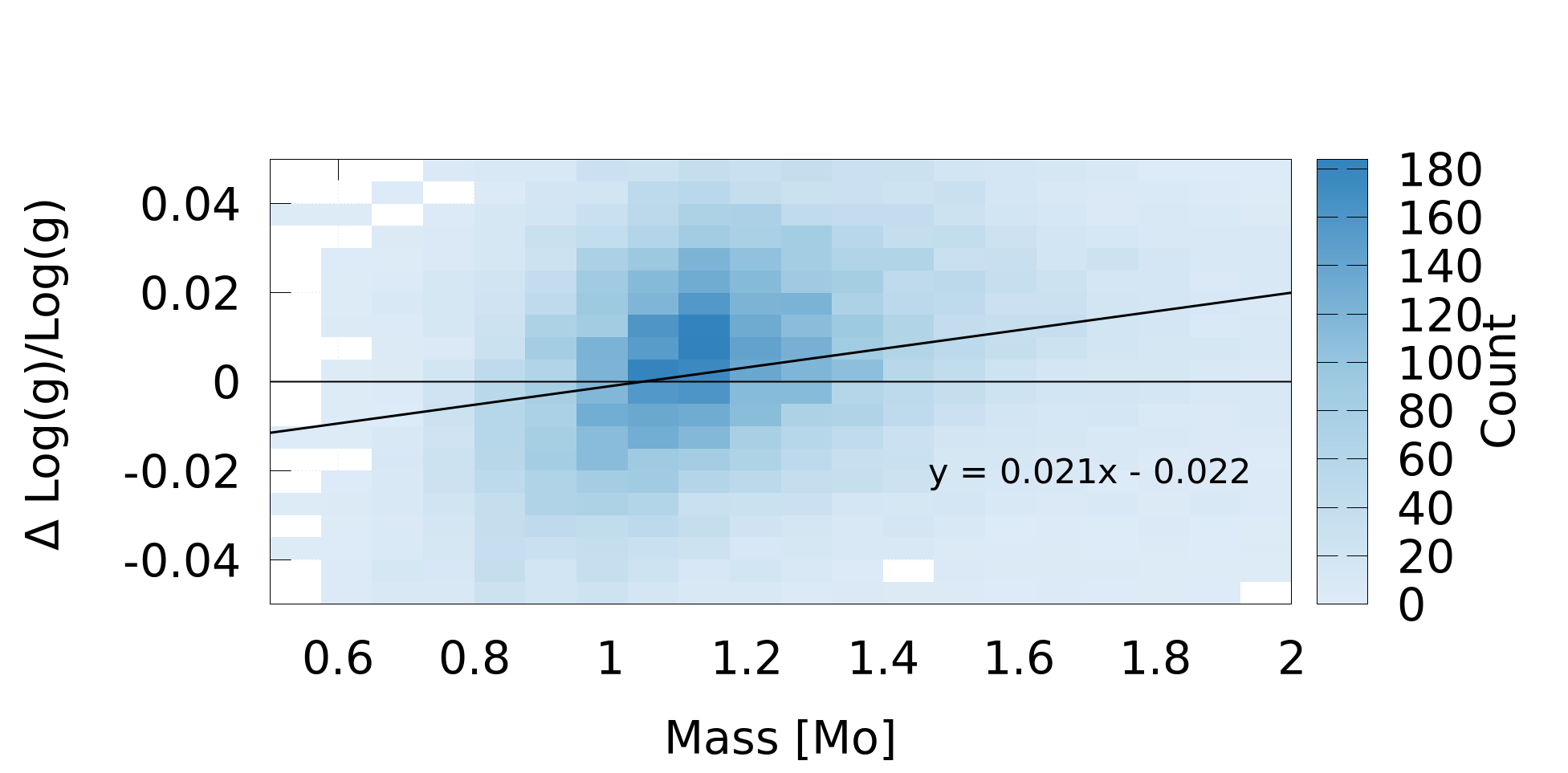}}
\subfigure{\includegraphics[width=8.5cm,clip=true, trim=0.0in 0.0in 0in 0.4in]{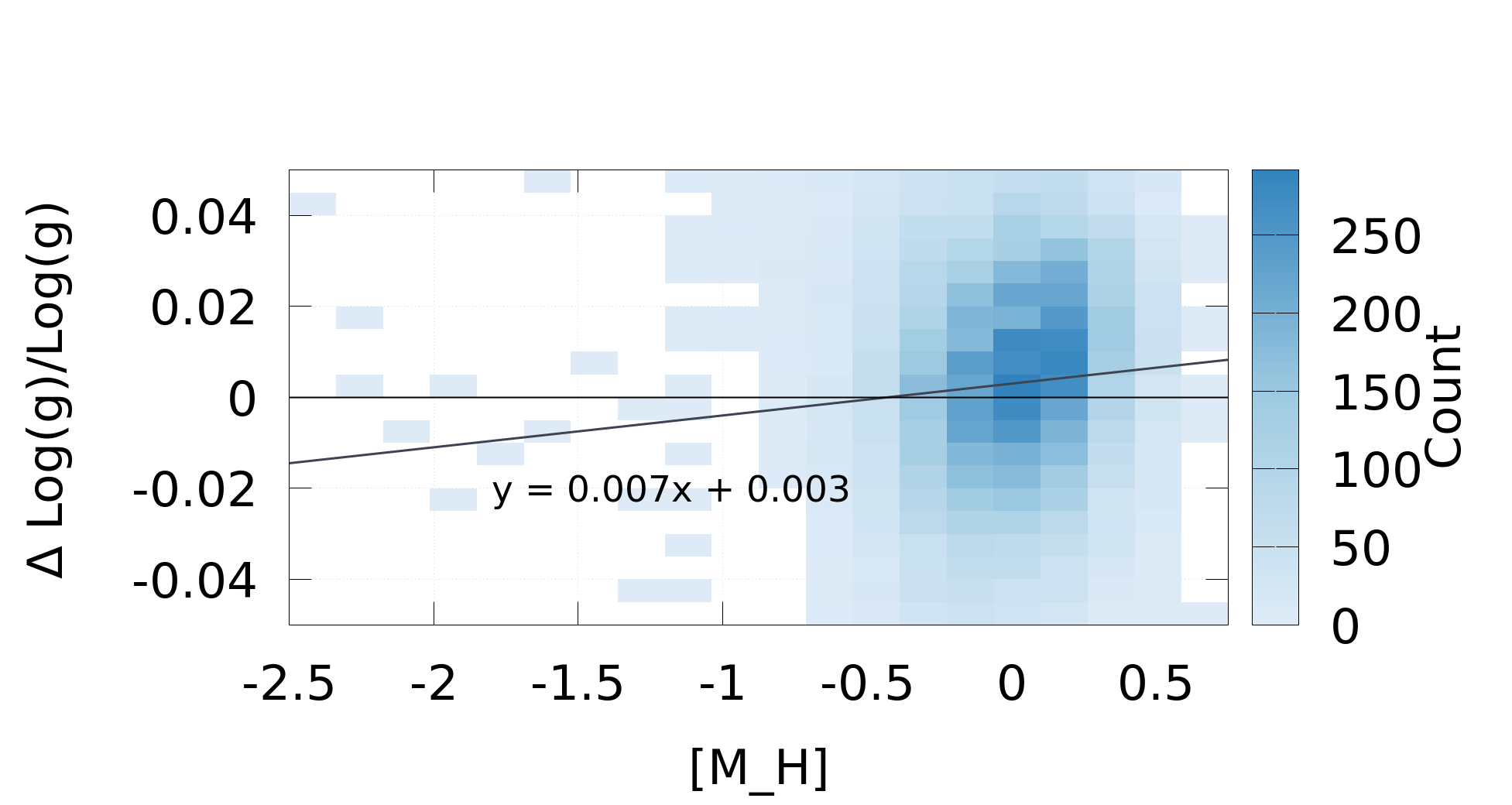}}
\subfigure{\includegraphics[width=8.5cm,clip=true, trim=0.0in 0.0in 0in 0.4in]{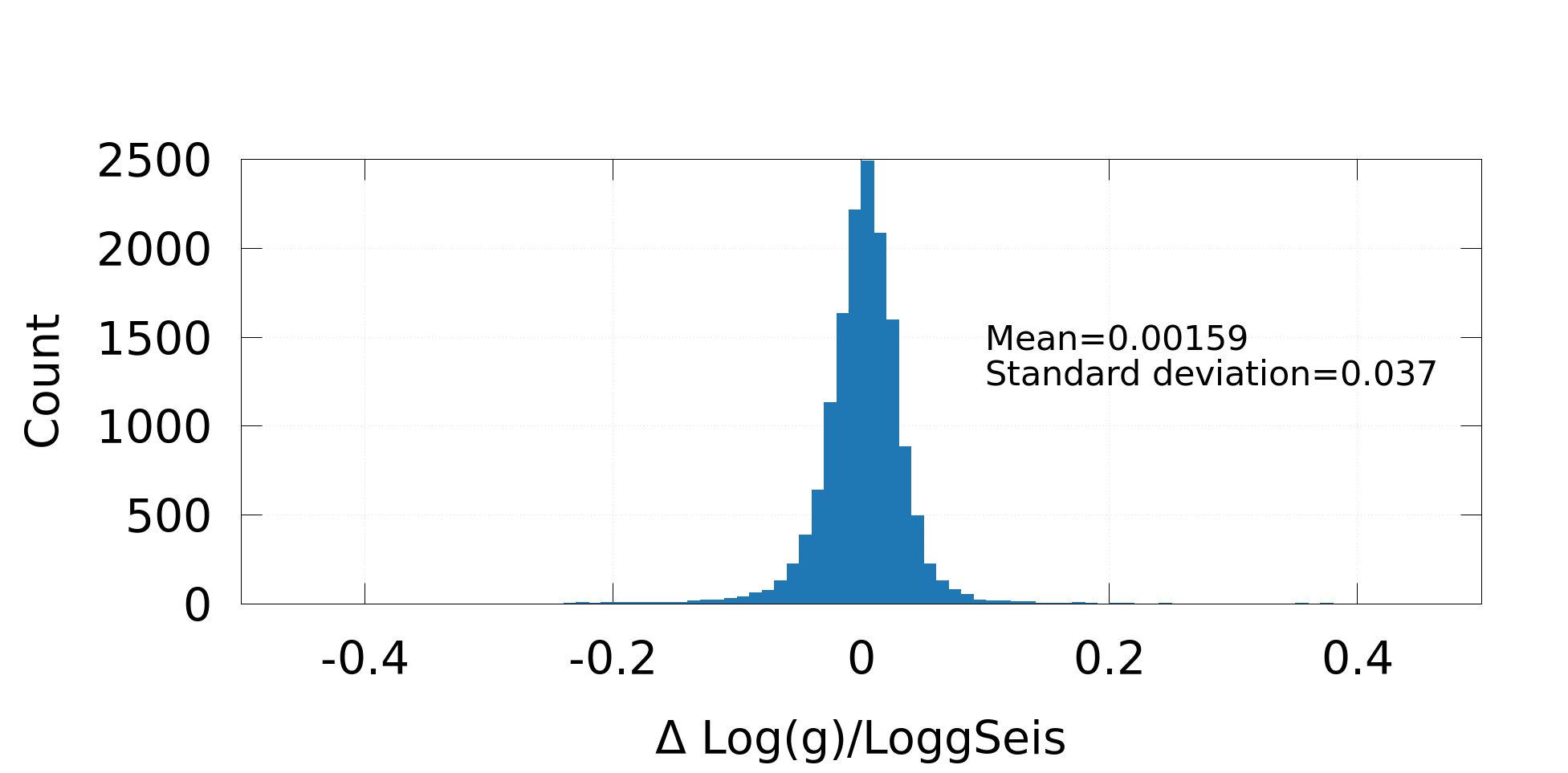}}
\subfigure{\includegraphics[width=8.5cm,clip=true, trim=0.0in 0.0in 0in 0.4in]{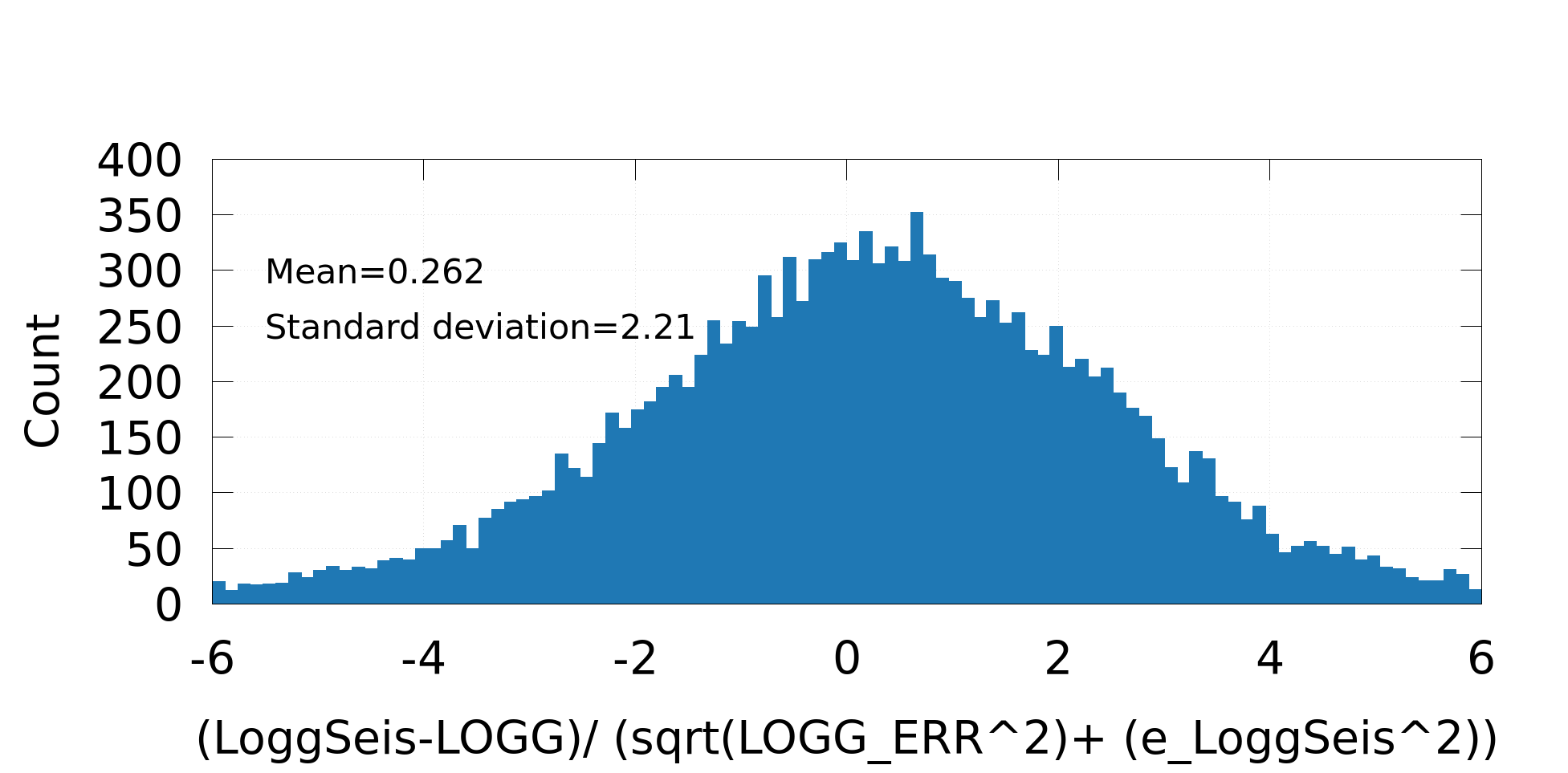}}
\caption{\textbf{Top four}: Offset between the spectroscopic and asteroseismic surface gravities (Seismic - Spectroscopic) compared to spectroscopic log(g), [$\alpha$/M], mass, and [M/H]. The darkness is representative of the concentration of stars in a given area; slight trends can be seen with mass and metallicity and in the red clump (top right). \textbf{Bottom left}: Offset between the spectroscopic and asteroseismic results. A mean of 0.001 and uncertainty of 3.5\% indicate excellent agreement between the measurements. \textbf{Bottom right}: Histogram comparing the measured offsets before the seismic and spectroscopic results scaled by the claimed uncertainties for each measurement. The distribution is much wider than the standard-normal distribution, suggesting that the claimed uncertainties are underestimated by a factor of 2.}
\label{Fig:NoTrends}
\end{center}
\end{figure}

\section{Analysis} \label{sec:Analysis}
Here we rely on the transitive property. We know that the APOGEE spectroscopic surface gravity was calibrated on the Kepler asteroseismic gravity scale. Therefore, we assume that if the TESS surface gravity matches the surface gravity from APOGEE, they must be consistent with the Kepler scale. In general, we found this to be the case; there is agreement to better than 5\%, signifying a very accurate calibration.
 We searched for possible dependencies of the offset between the asteroseismic and spectroscopic surface gravities on metallicity, [$\alpha/$M], log(g), temperature, and mass. (Figure \ref{Fig:NoTrends}). We found evidence for slight potential trends with mass, metallicity, and surface gravity. Previous work has identified significant inconsistencies in the asteroseismic results in surface gravity as a function of metallicity \citep{Epstein2014}; here, we see at most a slight trend. We note that the largest offsets that we discovered tended to be for the clump stars. In Figure \ref{Fig:NoTrends}, we show histograms of the difference between the TESS asteroseismic and APOGEE spectroscopic gravities scaled by the claimed uncertainty for each star. Given that this is significantly wider than the standard normal distribution, we assert that the predicted uncertainty is underestimated for the majority of stars. Comparisons between the Kepler and APOGEE data indicate that the spectroscopic uncertainties in DR17 may be underestimated by at least a factor of two (M. Pinsonneault et al., in prep.), but this would not be sufficient to match the observed differences in our sample, suggesting that there are additional uncertainties that are also underestimated. 

\section{Conclusion} \label{sec:Conclusion}
Our analysis indicates that the asteroseismic results from \citet{Hon2021} are reliable. For more than 98\% of stars, the inferred asteroseismic gravity matches the spectroscopic gravities to better than 10\%, suggesting that true giants have been identified. The precision for $\nu_{\rm max}$ is also good -- for 90\% of stars, the agreement on the inferred log(g) is better than 5\% ($\sim 0.05$ dex in log(g)). We observe slight differences between the spectroscopic and asteroseismic gravities that correlate with mass, log(g), and [M/H], but no significant offsets. We do note that there is a slightly larger offset on average for clump stars, which we attribute to one of the following:
1. A slight failure rate in correction identifying the stars' evolutionary state using only spectroscopic information, which is required to precisely calibrate the spectra to the Kepler asteroseismic scale.
2. Slight evolutionary state-dependent errors in the correction factor applied to the APOGEE spectroscopic gravities
3. Evolutionary state-dependent challenges in correctly pinpointing $\nu_ {\rm max}$ in short data sets using only a neural network. 
Given these results, we encourage both future investigations into the precise details of TESS asteroseismology of red giants, as well as usage of the current TESS results for galactic archaeology purposes.

\begin{acknowledgements} \label{sec:Acknowledgements}

NASA grants 80NSSC23K0143 and 80NSSC23K0436. We utilize data from SDSS 

(https://www.sdss.org/collaboration/citing-sdss/)

A.T Thanks Georgios and Hatice Theodoridis for helpful discussions
\end{acknowledgements}

\bibliography{paper.bbl}{}
\bibliographystyle{aasjournal}

\end{document}